\documentclass{kluwer}
\usepackage{graphicx}
\newcommand{\ion}[2]{\makebox{#1\,\textsc{#2}}}
\newcommand{\thd}{\ensuremath{\theta_d}}
\newcommand{\smunits}{\mbox{kpc~m${}^{-20/3}$}}
\newcommand{\kms}{\mbox{km~s${}^{-1}$}}
\newcommand{\cd}{\mbox{cm${}^{-2}$}}

\newcommand{\aap}{\textrm{A\&A}}
\newcommand{\apj}{\textrm{ApJ}}

\newcommand{\nat}{\textrm{Nature}}
\begin{document}
\begin{article}
\begin{opening}

\title{Extreme Scattering Events: An Observational Summary}
\author{T.~Joseph~W. \surname{Lazio}\email{lazio@rsd.nrl.navy.mil}}
	\institute{Naval Research Laboratory}
\author{Alan~L. \surname{Fey}\email{afey@usno.navy.mil}}
	\institute{US Naval Observatory}
\author{R.~A. \surname{Gaume}\email{rgaume@usno.navy.mil}}
	\institute{US Naval Observatory}

\runningtitle{ESEs: An Observational Summary}
\runningauthor{Lazio et al.}

\begin{abstract}
We review observational constraints on the structures responsible for
extreme scattering events, focussing on a series of observations of
the quasar PKS 1741$-$038.  VLA observations were conducted to search
for changes in the rotation measure and \ion{H}{i} absorption during
the ESE, while VLBI observations sought ESE-induced changes in the
source's image.  No RM changes were found implying $B_{||} < 12$~mG,
and no \ion{H}{i} opacity changes were found implying $N(\ion{H}{i}) <
6.4 \times 10^{17}$~\cd.  No multiple imaging was observed, but the
diameter of the source \emph{increased} by~0.7~mas, contrary to what
is predicted by simple refractive lens modeling of ESEs.  We summarize
what these limits imply about the structure responsible for this
\hbox{ESE}.
\end{abstract}
\end{opening}

\vspace*{-0.75cm}
\section{Introduction}

\begin{figure}
\centerline{\includegraphics[width=0.67\textwidth,angle=-90]{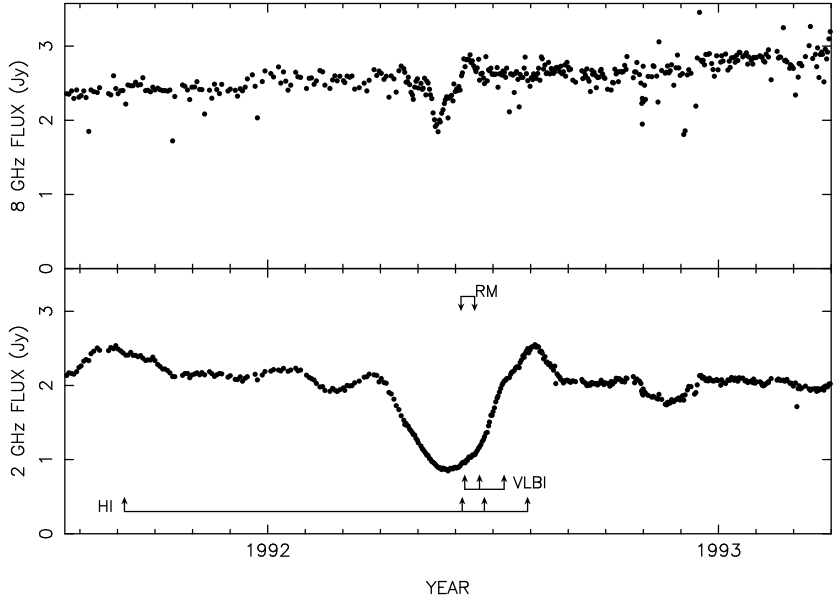}}
\caption[]{The light curves of 1741$-$038 at~2 and~8~GHz showing the
extreme scattering event and the epochs of the various observations.}
\label{fig:lightcurve}
\end{figure}

Extreme scattering events (ESE) are a class of dramatic decreases
($\gtrsim$ 50\%) in the flux density of radio sources near~1~GHz for
several weeks bracketed by substantial increases (\opencite{fdjws94};
Fig.~\ref{fig:lightcurve}).  Because of their simultaneity at
different wavelengths and light travel time arguments, ESEs are likely
a propagation effect \cite{fdjh87a}.  First identified toward
extragalactic sources, ESEs have since been observed toward pulsars
\cite{cblbadd93,mlc98}.

To date, the only other observational constraints on the structures
responsible for ESEs---besides the light curves---are the lack of
pulse broadening and the variation in the pulse times of arrival
during the pulsar ESEs.  This paper summarizes constraints obtained
during the ESE toward the quasar 1741$-$038.  We discuss Faraday
rotation measurements by \inlinecite{cff96} in \S\ref{lazio:sec:rm},
VLBI imaging by \inlinecite{lazioetal00a} in \S\ref{lazio:sec:vlbi},
and \ion{H}{i} absorption measurements by \inlinecite{lazioetal00b} in 
\S\ref{lazio:sec:hi}.  We present our conclusions in
\S\ref{lazio:sec:conclude}.  Figure~\ref{fig:lightcurve} shows the ESE 
of 1741$-$038 with the epochs of the various observations indicated.

\section{Faraday Rotation Measure Observations}\label{lazio:sec:rm}

At each epoch, the polarization position angle~$\phi$ was measured
at~6 to~10 frequencies and then fit, by minimizing $\chi^2$ and
accounting for $n\pi$ ambiguities, as a function of the observing
wavelength, $\lambda^2$.  The same procedure was used for 1741$-$038
and for the parallactic angle calibrator 1725$+$044.

Based on the relative invariability of RM toward 1725$+$044 and
1741$-$038, \inlinecite{cff96} conclude $\Delta\mathrm{RM} <
10.1$~rad~m${}^{-2}$ during the \hbox{ESE}.  This upper limit implies
an upper limit on the mean magnetic field parallel to the line of
sight of $B_{||} < 12$~\hbox{mG} for a typical value of the free
electron column density through the structure, $N_0 \sim
10^{-4}$~pc~cm${}^{-3}$ \cite{cfl98}.

This upper limit is much larger than the typical interstellar field
strength but enhancement of the ambient field to $B \sim 1$~mG is
possible within a shock front \cite{ccc88}.  Alternately,
$\Delta\mathrm{RM}$ may be small if the field is disordered or if the
ionized region is not magnetized.

\section{VLBI Imaging}\label{lazio:sec:vlbi}

Comparison of the visibility data during the ESE to those obtained
after the ESE (1994 July~8) show the source to be more resolved during
the \hbox{ESE}.  The excess angular broadening is $\Delta\thd \lesssim
0.7$~mas, implying that the ESE structure contributed a scattering
measure~$\mathrm{SM}_{\mathrm{ESE}} = 10^{-2.5}$~\smunits.  In turn,
the pulse broadening of a background pulsar should be $\tau_d \le
1.1D_{\mathrm{kpc}}$~$\mu$s at~1~GHz, consistent with the observed
lack of broadening during pulsar ESEs \cite{cblbadd93,mlc98}.  The
refractive models commonly used to explain ESEs predict that the
source's flux density and angular diameter should be highly
correlated.  We observe an \emph{anti-correlation}.  Simple stochastic
broadening models require much more scattering (2~mas) than is
observed.  We consider it likely that both refractive defocussing and
stochastic broadening are occurring.

We were unable to test a key prediction of refractive models---angular
position wander of the background source---because these observations
had no absolute position information.  A second prediction is multiple
imaging.  During this ESE, any secondary image(s) must have been
extremely faint; multiple imaging, with the secondary image slightly
offset from the primary, is unlikely to explain the increase in
angular diameter because no other effects of strong refraction are
seen (cf.\ also \opencite{cfl98}).  We also observe little, if any,
ESE-induced anisotropy in the VLBI images.  If ESE lenses are
filamentary structures \cite{rbc87}, they must be extended
\emph{along} the line of sight, a possibility also suggested by
\inlinecite{lrc98}.

\section{H\,I Absorption}\label{lazio:sec:hi}

At all epochs the \ion{H}{i} opacity spectra show the presence of a
strong absorption feature near~5~\kms\ and a typical rms
determined outside the \ion{H}{i} line of $\sigma_\tau \approx 0.015$.
There is no gross change in the absorption line during the ESE nor do
any additional absorption components appear.  Between any two epochs
$\Delta\tau \le 0.049$ ($< 2.3\sigma_\tau$).  This upper limit implies
a neutral column density change of $\Delta N_H < 6.4 \times
10^{17}\,\cd\,(T_s/10\,\mathrm{K})$ for a structure with a spin
temperature $T_s = 10$~\hbox{K}.

\inlinecite{h97} proposes interstellar tiny (AU)-scale atomic
structures (TSAS) in order to explain small (angular) scale changes in 
\ion{H}{i} opacity.  TSAS would have  $N_H \sim 3 \times 10^{18}$~\cd\
and $T_s \sim 15$~K.  Our $\Delta\tau$ limit marginally excludes a
connection between TSAS and ESEs.

\inlinecite{ww98} propose AU-scale molecular clouds in the Galactic
halo precisely to explain ESEs.  The clouds would be cold, $T_s
\gtrsim 3$~K, and dense enough to be opaque in the \ion{H}{i} line.
ESEs would result from the photoionized skins of the clouds.  We see
no $\tau \sim 1$ features (Walker~2000, private communication, has
since suggested $\tau \sim 0.1$).  However, the clouds could have
velocities approaching 500~\kms, while the observed velocity range is
no more than 250~\kms---significant \ion{H}{i} absorption could have
been present outside of our velocity range.

\section{Summary}\label{lazio:sec:conclude}

Salient aspects of this observational program are
\begin{itemize}
\item $\Delta\mathrm{RM} < 10.1$~rad~m${}^{-2}$ implying a magnetic
field within the scatterer of $B_{||} < 12$~\hbox{mG}.

\item No change in the VLBI structure, except a 0.7~mas
\emph{increase} in the angular diameter.  This increase is \emph{not}
consistent with that expected from a purely refractive model: ESEs
must result from both broadening and defocussing within the ionized
structures.

\item $\Delta\tau_{\mathrm{H\,I}} < 0.05$ implying that the \ion{H}{i}
column density associated with the ESE structure is $N_H < 6.4 \times
10^{17}\,\cd$.  Tiny-scale atomic structures are marginally ruled out;
\ion{H}{i}-opaque, halo molecular clouds would be excluded, but
the observed velocity range covers only 25\% of the allowed range.
\end{itemize}

The major impediment to improved observational constraints is the lack
of a monitoring program that could find additional ESEs.

\acknowledgements
Many people helped make these observations possible, most notably
M.~Claussen, A.~Clegg, B.~Dennison, R.~Fielder, K.~Johnston, and
E.~Waltman.  The NRAO is a facility of the National Science Foundation
operated by \hbox{AUI}.  Radio astronomy research at the NRL is
supported by the Office of Naval Research.

\end{article}
\end{document}